\documentclass[twocolumn,apj,numberedappendix]{templates/openjournal/openjournal}
\usepackage[utf8]{inputenc}

\usepackage{graphicx}	
\usepackage{amsmath}	
\usepackage{amssymb}	
\usepackage{bm}		
\usepackage{natbib}
\usepackage{float}
\usepackage{algorithm}
\usepackage[noend]{algpseudocode}
\usepackage{setspace}

\usepackage[dvipsnames]{xcolor}
\usepackage[most]{tcolorbox}
\usepackage{hyperref}
\hypersetup{
    colorlinks=true,
    linkcolor=blue,
    filecolor=blue,
    urlcolor=blue,
    citecolor=blue,
    }
\usepackage{fontawesome}
\makeatletter
\newcommand{\github}[1]{%
   \href{#1}{\faGithub}%
}
\makeatother

\providecommand{\norm}[1]{\lVert#1\rVert}
\DeclareMathOperator*{\argminT}{argmin}
\newcommand{\gmu}{\ensuremath{G \mu}}

\newcommand{\eg}{\mbox{\textit{e.g.}}}

\begin{document}

\journalinfo{The Open Journal of Astrophysics}
\submitted{submitted XXX; accepted YYY}

\shorttitle{Emulation of cosmic string signatures}
\shortauthors{Price, Mars, Docherty, Spurio Mancini, Marignier \& McEwen}

\title{Fast emulation of anisotropies induced in the cosmic microwave background by cosmic strings}

\author{M.~A.~Price$^{\dagger1}$}
\author{M.~Mars$^{1}$}
\author{M.~M.~Docherty$^{1}$}
\author{A.~Spurio Mancini$^{1}$}
\author{A.~Marignier$^{1,2}$}
\author{J.~D.~McEwen$^{1,3}$}

\affiliation{$^{1}$Mullard Space Science Laboratory (MSSL), University College London (UCL), Holmbury St Mary, Dorking, Surrey RH5 6NT, UK}
\affiliation{$^{2}$Department of Earth Sciences, UCL, London, UK}
\affiliation{$^{3}$ Alan Turing Institute, Euston Road, London NW1 2DB, UK}

\thanks{$^\dagger$ E-mail: \nolinkurl{m.price.17@ucl.ac.uk}}

\begin{abstract}
  Cosmic strings are linear topological defects that may have been produced during symmetry-breaking phase transitions in the very early Universe. In an expanding Universe the existence of causally separate regions prevents such symmetries from being broken uniformly, with a network of cosmic string inevitably forming as a result. To faithfully generate observables of such processes requires computationally expensive numerical simulations, which prohibits many types of analyses. We propose a technique to instead rapidly \emph{emulate} observables, thus circumventing simulation. Emulation is a form of generative modelling, often built upon a machine learning backbone. End-to-end emulation often fails due to high dimensionality and insufficient training data. Consequently, it is common to instead emulate a latent representation from which observables may readily be synthesised. Wavelet phase harmonics are an excellent latent representations for cosmological fields, both as a summary statistic and for emulation, since they do not require training and are highly sensitive to non-Gaussian information. Leveraging wavelet phase harmonics as a latent representation, we develop techniques to emulate string induced CMB anisotropies over a $7.2^{\circ}$ field of view, with sub-arcminute resolution, in under a minute on a single GPU. Beyond generating high fidelity emulations, we provide a technique to ensure these observables are distributed correctly, providing a more representative ensemble of samples. The statistics of our emulations are commensurate with those calculated on comprehensive Nambu-Goto simulations. Our findings indicate these fast emulation approaches may be suitable for wide use in, \emph{e.g.}, simulation based inference pipelines. We make our code available to the community so that researchers may rapidly emulate cosmic string induced CMB anisotropies for their own analysis. \github{https://github.com/astro-informatics/stringgen}
\end{abstract}


\maketitle

\section{Introduction}
\label{sec:introduction}

Cosmic strings are linear topological defects produced when the Universe undergoes certain symmetry-breaking phase transitions, arising for example in a range of attempts at Grand Unification; for reviews see \citealt{brandenberger:1994,vilenkin:1994,hindmarsh:1995,copeland:2009}. In an expanding Universe, the existence of causally separate regions prevents the symmetry from being broken in the same way throughout space, with a network of cosmic strings inevitably forming as a result~\citep{kibble:1976}.  Cosmic strings are thus a well-motivated extension of the standard cosmological model and, while a string network cannot be solely responsible for the observed anisotropies of the cosmic microwave background (CMB) (since they could not explain the acoustic peaks of the CMB power spectrum; \citealt{pen:1997}), they could induce an important subdominant contribution.

The amplitude of any CMB anisotropies induced by cosmic strings is related to the string tension $\gmu$, where $G$ is Newton's gravitational constant and $\mu$ is the energy per unit length of the string.  In turn, the energy scale $\eta$ of the string-inducing phase transition is directly related to $\mu$ by $\mu \sim \eta^2$. Detecting signatures of cosmic strings would therefore provide a direct probe of physics of phase transitions in the early Universe at extremely high energy scales.  Consequently, there has been a great deal of interest in constraining cosmic strings using observations of the CMB.  In the majority of such analyses, signatures of string observables must be simulated, which is highly challenging.

Simulating accurate observable effects of a network of cosmic strings is a rich and highly computationally demanding field of research \citep{albrecht:1989, bennett:1989, bennett:1990, allen:1990, hindmarsh:1993, bouchet:1988, vincent:1998, moore:2001, landriau:2002, ringeval:2005, fraisse:2007, landriau:2010, blancopillado:2011, ringeval:2012}.  There is an ongoing disagreement between Nambu-Goto \citep[\eg][]{ringeval:2012} and Abelian Higgs \citep[\eg][]{hindmarsh:2017} simulation models regarding the decay of loops in string networks.  In any case, in both models large-scale numerical simulations are required to faithfully evolve string networks and simulate their observational effects. For example, the simulation of a single full-sky Nambu-Goto string-induced CMB map at sub-arcminute angular resolution can require in excess of 800,000 CPU hours, which is only possible by massively parallel ray tracing through thousands of Nambu-Goto string simulations \citep{ringeval:2012}.

A variety of methods have been developed to search for string-induced contributions to the CMB, including power-spectrum constraints~\citep{lizarraga:2014,lizarraga:2014b,lizarraga:2016,charnock:2016}, higher-order statistics such as the bispectrum~\citep{planck2013-p20,regan:2015} and trispectrum~\citep{fergusson:2010}, and approaches such as edge detection~\citep{lo:2005,amsel:2008,stewart:2009,danos:2010}, Minkowski functionals~\citep{gott:1990,ducout:2013}, wavelets and curvelets~\citep{starck:2004,hammond:2009,wiaux:2010:csstring,planck2013-p20,hergt:2017,mcewen:cosmic_strings_swbd}, level crossings~\citep{movahed:2011}, peak-peak correlations~\citep{movahed:2013} and Bayesian inference~\citep{mcewen:cosmic_strings_swbd, ciuca:2017}. More recently, machine learning techniques have also been developed and shown great effectiveness \citep{ciuca:2019:convolutional, ciuca:2019:inferring, ciuca:2020, vafaei:2018, torki:2022}.
Due to the discrepancies between string simulation models, current constraints on the string tension depend on the model and simulation technique adopted.  We avoid surveying the various constraints that have been reported in the literature to date and merely remark that typical constraints bound the string tension by $\gmu \lesssim 10^{-7}$ \citep[\eg][]{planck2013-p20}.

A critical component of all approaches to search for a cosmic string contribution in the CMB is the accurate simulation of string-induced CMB anisotropies. The massive computational cost of accurate string simulations, irrespective of the string simulation model, limits the effectiveness of cosmic string searches. This massive computational cost is currently unavoidable if the string network is to be accurately evolved and observables simulated faithfully. Compounding this, since strings induce significant contributions to CMB anisotropies at small angular scales, observables must be simulated at high-resolution.
These limitations motivate alternative machine learning-based \textit{emulation} techniques to generate realisations of synthetic observables, without the prohibitive computational overhead of full physical simulations, which is the focus of this article.
Emulation is closely related to generative modelling and borrows many of the core ideas; naturally, many emulation methods leverage modern machine learning models, \textit{e.g.} variational auto-encoders \citep{kingma:2013}.

While techniques to emulate cosmic string-induced CMB anisotropies accurately do not exist currently, as far as we are aware, approaches to emulate other cosmological fields, such as large-scale structure, have been considered. Generative adversarial networks \citep{rodriguez:2018, mustafa:2019, perraudin:2021:emulation, feder:2020} and variational auto-encoders \citep{chardin:2019:deep} have found some success emulating density fields directly \citep{piras:2023}. However, such end-to-end approaches are limited to low to moderate dimensions and require large volumes of training data.  To circumvent the issues of high dimensionality and large volumes of training data, an alternative approach is to emulate some latent representation from which observables may be readily synthesised. For example, it is common to first emulate a power spectrum, \emph{e.g} through polynomial regression \citep{jimeneze:2004, fendt:2007:pico}, Gaussian processes \citep{heitmann:2009:coyote,lawrence:2010:coyote,ramachandra:2021,euclid:2021}, or multilayer perceptrons \citep{auld:2008:cosmonet, agarwal:2012:pkann,bevins:2021:globalemu,spuriomancini:2022}, from which Gaussian realisations may trivially be generated. For the emulation of string-induced anisotropies, which are highly non-Gaussian, adopting the power spectrum as a latent representation is not well-suited.

In this article we propose a technique to emulate CMB anisotropies induced by networks of cosmic strings that both eliminates the computational bottleneck and captures non-Gaussian structure. Our emulation technique adopts the recently developed wavelet phase harmonics \citep{mallat:2020:phase,allys:2020:new,zhang:2021:maximum, brochard:2022}, a form of second generation scattering transform \citep{mallat:2012, bruna:2013}, as a latent representation. Once a wavelet phase harmonic representation is computed from a \emph{small ensemble} of physical simulations, our approach can then be used to rapidly generate high-resolution realisations of the cosmic string induced CMB anisotropies in under a minute, starkly contrasting the computational cost of a single simulation. Such an acceleration unlocks a variety of analysis techniques, including but not limited to those which necessitate the repeated synthesis of observables, \emph{e.g.} Bayesian inference which often relies on sampling. In particular our approach is suitable for use in simulation based inference (SBI) pipelines \citep{cranmer:2020, mancini:2022:bayesian}, where the likelihood is either not available or too costly to be evaluated, and inference relies solely on the ability to efficiently simulate or emulate observables. Such techniques are predicated on the ability to generate observations that are not only realistic but are also correctly distributed. In this article we explore this second qualification as well, which is often overlooked despite being critical for scientific studies.

The remainder of this article is structured as follows. In Section \ref{sec:generative_modelling} we provide an overview of generative modelling within the context of cosmology. We then present our approach for the rapid emulation of cosmic string induced CMB anisotropies in Section \ref{sec:fast_emulation}, which we subsequently validate in Section \ref{sec:numerical_results}. Finally, we discuss the impact of these results and draw conclusions in Section \ref{sec:conclusions}.

\section{Generative Modelling of Physical Fields}
\label{sec:generative_modelling}
Generative modelling is a term broadly ascribed to the generation of synthetic observables that approximate authentic observables. Throughout the following discussion we will refer to authentic observables by $\bm{x}_{\text{\tiny True}}$ and synthetic observables by $\bm{x}_{\text{\tiny Syn}}$, which can be either simulated or emulated observables, denoted $\bm{x}_{\text{\tiny Sim}}$ and $\bm{x}_{\text{\tiny Emu}}$ respectively. A diverse range of generative models exist with varying motivations, although many are motivated by the \emph{manifold hypothesis} \citep{bengio:2013:representation}.

\vspace{7pt}
\noindent \textbf{Manifold Hypothesis:}~\textit{A given authentic observable $\bm{x}_{\text{\tiny True}} \in \mathcal{X}$, where $\mathcal{X}$ is the \emph{ambient} space with dimensionality $d_{\mathcal{X}}$, is hypothesised to live on a manifold $\mathcal{S} \subseteq \mathcal{X}$ with dimensionality $d_{\mathcal{S}} \leq d_{\mathcal{X}}$, embedded within $\mathcal{X}$.}
\vspace{7pt}

Intuitively, this becomes apparent by considering natural images and making the following realisations. First, images generated by uniformly randomly sampling each pixel are extremely unlikely to be meaningful \citep{pope:2021:intrinsic}. Secondly, images are highly locally connected through various transformations (\emph{e.g.} contrast, brightness), symmetries (\emph{e.g.} translations, scaling), and diffeomorphisms (one-to-one invertible mappings, \emph{e.g.} stretching). There is strong evidence to suggest the manifold hypothesis is correct \citep{bengio:2013:representation}, with algorithmic verification by \citet{fefferman:2016}. In any case, where additional flexibility is necessary a union of manifolds hypothesis may be adopted with similar justification \citep{brown:2022}.

For a complete description of the generative model one must also characterise the data generating distribution on this manifold, \emph{i.e.} the likelihood with which any given synthetic observable is to have been observed. In such a case one may interpret $\mathcal{S}$ as a \emph{statistical manifold} \citep[see \emph{e.g.}][]{amari:2016,nielsen:2020}.

\vspace{6pt}
\noindent \textbf{Statistical Manifold:}~\textit{A manifold $\mathcal{S}$ on which observables $\bm{x}_{\text{\tiny True}} \in \mathcal{S}$ live that is endowed with a probability distribution $\mathbb{P}_{\text{\tiny True}}$.}
\vspace{6pt}

Under the statistical manifold hypothesis the generative problem is two-fold: (i) how best to generate realistic synthetic observables $\bm{x}_{\text{\tiny Syn}} \in \mathcal{S}$, and (ii) how to ensure the probability distribution $\mathbb{P}_{\text{\tiny Syn}}$ of $\bm{x}_{\text{\tiny Syn}}$ matches $\mathbb{P}_{\text{\tiny True}}$. That is, how best to not only approximate the embedded manifold but also the distribution over that manifold. With machine learning techniques problem (i) can often be satisfied, provided access to a sufficiently large bank of data $\bm{d}$. However problem (ii) is less straightforward to address and in many cases depends on the degree to which the distribution of $\bm{d}$ traces $\mathbb{P}_{\text{\tiny True}}$.
It should be noted that, attempting to model both $\mathcal{S}$ and $\mathbb{P}_{\text{\tiny True}}$ with maximum-likelihood based methods can be pathological when the ambient dimensionality of the space $\mathcal{X}$ is significantly different to that of $\mathcal{S}$ \citep{dai:2018:diagnosing}. At a high-level this effect, which is referred to as \emph{manifold overfitting}, occurs when the manifold $\mathcal{S}$ is learned but the distribution over $\mathcal{S}$ is not \citep{loaiza-ganem:2022}.

One way in which this pathology may be solved is by first learning the data-distribution on a latent representation (equivalently a summary statistic) with low dimensionality (ideally equal to that of $\mathcal{S}$) before decoding to an approximation of the data-distribution. This approach to learning the data distribution was first explored by \citet{loaiza-ganem:2022}, who show that if the latent representation is a generalized autoencoder, then the data-distribution on $\mathcal{S}$ may be recovered theoretically \citep[see][Theorem 2]{loaiza-ganem:2022}. A variety of other effective methods have been proposed to handle this pathology \citep{arjovsky:2017,horvat:2021,song:2019,song:2020}.

The importance of the above criteria when generating natural images or physical fields differs greatly. In most applications, it is sufficient to rapidly generate inexpensive synthetic observables with high fidelity. For example, in the large-scale generation of synthetic natural images or celebrity faces \cite{rombach:2022}, matching the correct data generating distribution $\mathbb{P}_{\text{\tiny True}}$ is perhaps less important. For scientific analysis, however it is typically necessary to generate synthetic observables that not only live on or in the neighbourhood of $\mathcal{S}$, but also are approximately drawn from $\mathbb{P}_{\text{\tiny True}}$. An accurate approximation of the distribution on the manifold is critical for use in, for example, simulation based inference pipelines.
\subsection{Simulation} \label{sec:sub_simulation}

Many generative models have been developed for a broad range of applications, however in this article we will consider two categories: simulation and emulation. From the perspective of a cosmologist, simulation entails the time evolution of initial conditions, \emph{e.g.} an initial field $\bm{x}_{0}$, governed by cosmological parameters $\bm{\theta}$, to some late-universe observables $\bm{x}$. Such evolution is designed to model the underlying physics of a universe from the grandest to smallest scales, which can become incredibly complex and non-linear \citep{hockney:2021}. Extracting information at higher angular resolutions is of increasing importance as recent and forthcoming cosmological experiments probe smaller scales with greater sensitivity. Simulating small-scale physics is therefore critical, necessitating high resolution simulations to faithfully represent late-universe observables, which is highly computationally demanding. Computationl hurdles aside, it is important to note that, provided the core physics is sufficiently captured, an ensemble of simulated observables will reliably trace $\mathbb{P}_{\text{\tiny True}}$, which is critical for subsequent analyses.

\vspace{6pt}
\noindent \textbf{Simulation:}~\textit{A generative model which directly encodes the dynamics of a physical system, evolving some initial conditions over time to a late universe observable $\bm{x}_{\text{\tiny Sim}}$. The dynamics of a system are governed by parameters $\bm{\theta}$.}
\vspace{-2pt}

Such generative models are dependent only on an understanding of both the initial conditions, parameters $\bm{\theta}$, and the underlying physics, and do not need to model the statistical distribution of the data directly since it is captured implicitly by the simulation process.

\subsection{Emulation} \label{sec:sub_emulation}

One may instead emulate observations, circumventing simulation entirely by approximating a mapping from cosmological parameters $\bm{\theta}$ to synthetic late-universe observables $\bm{x}_{\text{\tiny Emu}}$. Provided training data $\bm{d} = \lbrace \bm{\theta}, \bm{x}_{\text{\tiny True}} \rbrace$ one may attempt to train a model to approximate this mapping directly.
End-to-end approaches are reliant on a sufficiently large volume of training data, the amount of which scales with both dimensionality and functional complexity. Cosmology is fundamentally restricted to synthetic training data, which can only be accurately and reliably generated through computationally expensive simulations. While generating small numbers of such simulations is expensive but achievable \citep{nelson:2019, villaescusa:2020:quijote}, generating large ensembles of such simulations is often simply not feasible.

Consequently, to ameliorate these concerns it is common to instead emulate a compressed latent representation from which observables may readily be synthesised. In the following we define a compression $\Phi: \bm{x} \mapsto \bm{z} \in \mathcal{Z}$, where $\mathcal{Z}$ is of dimension $d_{\mathcal{Z}}$. Further, consider the setting where we constrain the ratio $ r = d_{\mathcal{Z}} / d_{\mathcal{X}} < 1$, such that $\bm{z}$ is a potentially lossy compressed representation of $\bm{x}$. The objective is therefore to first approximate the latent mapping $\bm{\Lambda}: \bm{\theta} \mapsto \bm{z}_{\text{\tiny Emu}}$ from which observables may be synthesised by taking into consideration the latent compression $\bm{z}_{\text{\tiny Emu}}=\Phi(\bm{x}_{\text{\tiny Emu}})$. To learn an approximation of $\bm{\Lambda}$ requires less training data due to the reduction in dimensionality. Hence, a trade-off between the complexity of $\bm{\Lambda}$ and $\Phi$ exists and so one can balance between data requirements and the information lost during compression. As the compression ratio $r$ decreases, \emph{i.e.} greater compression, the data requirements diminish, however conversely the compression loss is likely to increase.

Popular summary statistics such as the power-spectrum are emulated in this manner, from which Gaussian realisations may be generated trivially \citep[see \emph{e.g.}][]{auld:2008:cosmonet, agarwal:2012:pkann,bevins:2021:globalemu,spuriomancini:2022}. However, the power spectrum is a particularly ill-suited latent representation for the synthesis of cosmic string induced CMB anisotropies, which are highly non-Gaussian in nature. Hypothetically, one could adopt a variational auto-encoder \citep{kingma:2013} as an effective latent representation; in fact \citet{loaiza-ganem:2022} have recently had some success in this regard. It is reasonable to presume such an approach would be sensitive to non-Gaussian information, however for aforementioned reasons gathering sufficient training data is infeasible. This dichotomy therefore motivates the development of latent representations that are sensitive to non-Gaussian information and do not require substantial training data.

\vspace{6pt}
\noindent \textbf{Latent emulation:}~\textit{A two-step generative model, including a mapping $\bm{\Lambda}$ from cosmological parameters $\bm{\theta}$ to latent variables $\bm{z}_{\text{\tiny Emu}}$, from which observables $\bm{x}_{\text{\tiny Emu}}$ are synthesised given knowledge of the compression mapping $\Phi$ that maps from observables to the latent space, \emph{i.e.} $\bm{z}_{\text{\tiny Emu}} = \Phi(\bm{x}_{\text{\tiny Emu}})$.}
\vspace{6pt}

The reduced dimensionality of $\bm{z}_{\text{\tiny Emu}}$ alleviates training data requirements, introducing a trade-off between the complexity of the mapping and the compression loss, which can effect the quality of synthesis. Since $\Phi$ need only be surjective (onto), there typically exists some variability in synthetic observables, as potentially many observables correspond to a single latent vector. However, this implicit variability is by no means guaranteed to match the data generating distribution $\mathbb{P}_{\text{\tiny True}}$ on $\mathcal{S}$.
One should note that in the setting of \citet{loaiza-ganem:2022}, where $\Phi$ is a generalized autoencoder, provided $d_{\mathcal{Z}} = d_{\mathcal{S}}$, and the distribution on $\mathcal{Z}$ is sufficiently captured, the induced distribution $\mathbb{P}_{\text{\tiny Syn}}$ recovers $\mathbb{P}_{\text{\tiny True}}$ to a good approximation. Such an approach is appropriate for computer vision tasks, where data is far from a limiting factor. However, for cosmological applications insufficient data is available to learn such latent representations, motivating the adoption of designed representations, \emph{e.g.} wavelet-based representations.

\subsection{Wavelet Phase Harmonics}
The wavelet phase harmonics (WPH) are a form of second generation scattering transform \citep{mallat:2012,allys:2019:rwst,mallat:2020:phase} which can be directly contrasted with convolutional neural networks. For WPHs filters are defined by wavelets rather than learned in a data-driven manner.
Drawing inspiration again from machine learning, once a signal of interest has been convolved with the wavelet of a given scale, point-wise non-linearities are applied through the phase harmonic operator $\bm{w} \mapsto [\bm{w}]^p = |\bm{w}|\cdot e^{ip\text{arg}(\bm{w})}$, which is simply a rotation of some complex vector $\bm{w}$.
As such, rotations induce magnitude and scale independent non-linearities, hence spatial information may be synchronised across scales, from which moments (covariances between distinct convolutions) are computed. Consequently WPH provide a latent representation particularly well suited for spatially homogeneous images, \emph{e.g.} textures \citep{zhang:2021:maximum}. Furthermore, WPH can be shown to be highly sensitive to non-Gaussian information \citep{portilla:2000}, making them ideal latent representations for cosmic string induced CMB anisotropies.

WPH and their predecessors, the first generation wavelet scattering transform, have successfully been applied to probe weak gravitational lensing \citep{cheng:2020:new, cheng:2021:weak, valogiannis:2022, eickenberg:2022:wavelet}, the removal of non-Gaussian foreground contaminants \citep{allys:2019:rwst,blancard:2020:statistical,regaldo:2021:new,jeffrey:2022:single}, classifcation of magnetohydrodynamical simulations \citep{saydjari2021:classification}, and exploration of the epoch of reionisation \citep{greig:2022, lin:2022}. Many of these applications have adopted the WPH as a latent representation from which realistic observables have been emulated. However, as far as we are aware, to date little consideration has been given to the probability distribution of such observables (see Section \ref{sec:generative_modelling}).

There are two distinct way to construct maximum entropy generative models, these being the micro- and macro-canonical approaches, which relate to the associate ensembles in statistical physics. We have discussed the micro-canonical case, wherein new realizations which have the same latent representation are iteratively generated, and provided an arguement to why such an approach can result in limited variability. In contrast to this, the macro-canonical case consists in explicitly constructing a probability distribution for which the WPH are not fixed. This probability distribution can in turn be related to the physical Hamitonian of the process under study, however the difficulty is then how one samples from this ensemble \citep{marchand:2022:wavelet}.

\section{Fast Emulation of Cosmic String Induced Anisotropies} \label{sec:fast_emulation}
Having outlined both generative modelling and latent emulation in the context of physical fields, we next describe how these concepts can be leveraged to rapidly generate realisations of late-universe cosmic string induced anisotropies in the CMB.

\begin{figure*}
  \centering
  \includegraphics[width=\linewidth]{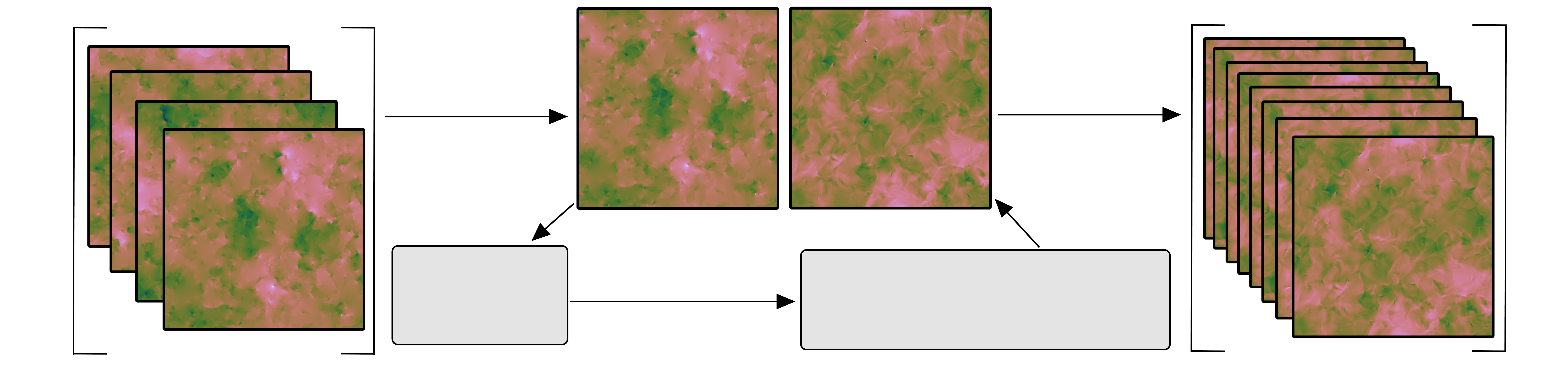}
  \put(-500,15){\rotatebox{90}{Simulation Ensemble}}
  \put(-16,107){\rotatebox{-90}{Emulation Ensemble}}
  \put(-295,110){$\sim 1$ day}
  \put(-216,110){$\sim 10$s}
  \put(-341,55){\raisebox{.0pt}{\textcircled{\raisebox{-.6pt} {1}}}}
  \put(-176,49){\raisebox{.0pt}{\textcircled{\raisebox{-.6pt} {2}}}}
  \put(-379,90){\scalebox{.85}{\scriptsize $\bm{x}_{\text{\tiny Sim}} \sim \lbrace \bm{x} \rbrace_{\text{\tiny Sim}}$}}
  \put(-182,90){\scalebox{.85}{\scriptsize $\bm{x}_{\text{\tiny Emu}} \rightarrow \gmu \bm{x}_{\text{\tiny Emu}}$}}
  \put(-312,30){\scalebox{.85}{\scriptsize $\bm{x}_0 \sim \mathcal{N}(0, \mathbb{I}_{d_{\mathcal{X}}})$}}
  \put(-377,27){\scalebox{.7}{$\bm{z}_{\text{\tiny Sim}} = \Phi(\bm{x}_{\text{\tiny Sim}})$}}
  \put(-380,16){\scalebox{.5}{Compute latent vector.}}
  \put(-243,25){\scalebox{.7}{$\bm{x}_{\text{\tiny Emu}} = \argminT \Big \lbrace \|\Phi(\bm{x}) - \bm{z}_{\text{\tiny Sim}}\|_2^2 \Big \rbrace$}}
  \put(-208,20){\tiny $\bm{x}$}
  \put(-233,13){\scalebox{.5}{Solve using automatic differentiation.}}
  \caption{An overview of the process by which a small ensemble of simulated observations can be extremely augmented with emulated observations for arbitrary string tension $\gmu$. In step 1 (compression) we simply draw a uniform random simulation $\bm{x}_{\text{\tiny Sim}}$ from which a reference latent vector $\bm{z}_{\text{\tiny Sim}}$ is calculated. In step 2 (synthesis) we take a random Gaussian realisation $\bm{x}_0$ and, using automatic differentiation of the compression mapping $\Phi$ to iteratively minimise a standard $\ell_2$-loss function, recovering solutions $\bm{x}_{\text{\tiny Emu}}$ such that $\Phi(\bm{x}_{\text{\tiny Emu}}) = \bm{z}_{\text{\tiny Sim}}$. Steps 1 and 2 can straightforwardly be repeated many times, generating an ensemble of emulated maps which can be (potentially much) larger than the small collection of simulated observables. In this way this approach may be thought of as extreme data augmentation.}
  \label{fig:synthesis}
  \vspace{7pt}
  \hrule
\end{figure*}
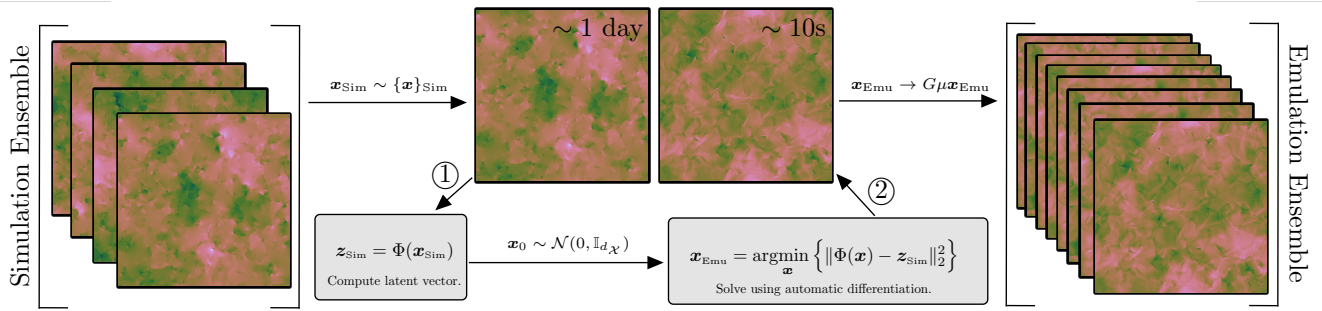

First let us explicitly formulate our emulation problem following the notation of Section \ref{sec:sub_emulation}. We seek to emulate late-universe string induced anisotropies $\bm{x}_{\text{\tiny Emu}}$ from cosmological parameters $\bm{\theta}$. Fortunately, in the case of cosmic string induced CMB anisotropies there is only a single parameter $\bm{\theta} = \gmu$, the string tension \citep{kibble:1976}. Moreover, the observed anisotropies transform trivially under $\mu \rightarrow \mu^\prime$, specifically this transformation is simply a scaling $\bm{x}_{\text{\tiny Emu}} \rightarrow (\mu^\prime/\mu) \bm{x}_{\text{\tiny Emu}}$. Therefore, provided one is able to generate emulated observables for a single string tension, it is straightforward to generate them for all string tensions. As such, in the following we simplify to a single fixed $\mu$ from which all $\mu^\prime\neq\mu$ can readily be generated \emph{a posteriori}.

We consider how to robustly synthesise string induced anisotropies $\bm{x}_{\text{\tiny Emu}}$ from their WPH representation $\bm{z}_{\text{\tiny Emu}}$. More formally, for a given reference latent vector $\bm{z}_{\text{\tiny Emu}}$, which we condition on, we efficiently synthesise observations $\bm{x}_{\text{\tiny Emu}}$ which satisfy $\Phi(\bm{x}_{\text{\tiny Emu}}) = \bm{z}_{\text{\tiny Emu}}$.
Additionally, we provide a strategy by which ensembles of such emulated observables can, at least approximately, be shown to be distributed appropriately, \emph{i.e.} $\mathbb{P}_{\text{\tiny Emu}} \approx \mathbb{P}_{\text{\tiny True}}$. To this end we leverage a small set of simulated observables $\bm{x}_{\text{\tiny Sim}}$ as a trellis, upon which our emulation process grows.

Throughout this work we will adopt WPHs as our compressed latent representation $\Phi$ \citep{mallat:2020:phase}, which is highly sensitive to non-Gaussian information \citep{portilla:2000}, is numerically efficient to evaluate, and does not require training data since it adopts designed rather than learned filters. We make use of the GPU-accelerated \texttt{PyTorch} package \texttt{PyWPH} \github{https://github.com/bregaldo/pywph}\footnote{\url{https://github.com/bregaldo/pywph}} which implements the transform discussed in \citet{regaldo:2021:new}, and by default adopts bump steerable wavelets \citep{mallat:2020:phase}.

\subsection{Generating String Induced Anisotropies} \label{sec:approximating_manifold}
In the following we work under the assumption that a (potentially very) limit number of simulated observables are available, from which we will generate arbitrarily many synthetic observables. Offline, we apply $\Phi$ to compress this \emph{training} set into latent vectors that we condition on during synthesis. We then iteratively emulate many observations $\bm{x}_{\text{\tiny Emu}}$ such that $\Phi(\bm{x}_{\text{\tiny Emu}}) \approx \bm{z}_{\text{\tiny Sim}}$ through gradient-based algorithms given an appropriate loss surface. Here we chose to minimise the standard Euclidean $\ell_2$-loss $\mathcal{L}(\bm{x}) = \norm{\Phi(\bm{x}) - \bm{z}_{\text{\tiny Sim}}}_2^2$. To achieve this in practice requires software to calculate both the compression $\Phi$ and necessary gradients, both of which are straightforwardly provided by \texttt{PyWPH}. An iterative approach, such as the one presented here, has also been adopted to successfully emulate a variety of cosmological signals, from density fields \citep{allys:2019:rwst,allys:2020:new} to foreground contaminants \citep{regaldo:2021:new, jeffrey:2022:single}.

For our current work we match the latent representation by maximum-likelihood estimation. One may instead perform maximum-a-posteriori estimation by enforcing regularity constraints. For example, cosmic string networks are close to piece-wise constant, hence emulation of their induced anisotropies may benefit from a total-variation norm regularisation (gradient sparsity), however we leave this exploration to a later date.

In this work we use the L-BFGS algorithm to minimise the loss function, which is a variant of the quasi-Newton method BFGS \citep{byrd:1995:limited} and typically require at most $100$ iterations to converge to a solution $\bm{x}_{\text{\tiny Emu}}$ for which the loss functions is below an acceptable tolerance. Visually, we confirm that these solutions $\bm{x}_{\text{\tiny Emu}}$ display similar characteristics to those generated through comprehensive simulations, indicating that they live on, or in the neighbourhood of, the embedded manifold $\mathcal{S}$. In many cases generating visually realistic synthetic observables alone is sufficient, \emph{e.g.} for natural images. However, to leverage these techniques for scientific inference it is important to ensure an ensemble of synthetic observables are distributed according to the data generating distribution $\mathbb{P}_{\text{\tiny True}}$.

\subsection{Matching the Probability Distribution} \label{sec:approximating_measure}
Suppose a single simulation is available, from which $m$ synthetic observables $\lbrace \bm{x}_{\text{\tiny Emu}} \rbrace_m$ may readily be emulated. From the surjectivity of $\Phi$ our emulated set of observables will exhibit some degree of variability, however this distribution is by no means guaranteed to match the true underlying data-generating distribution. In fact this is highly unlikely.

Were one to evaluate the expectation $\mathbb{E}[\cdot]$ of a summary statistic of interest $\Omega$ over these $m$ emulated observables they are likely to approximate the point statistics of the single simulation, but may not match the summary statistics averaged over an ensemble of $n$ simulations $\lbrace \bm{x}_{\text{\tiny Sim}} \rbrace_n$. This is to say that although our ensemble of emulated realisations sufficiently match a single simulation, they do not correctly characterise an ensemble of simulations $\mathbb{E}[\lbrace \Omega(\bm{x}_{\text{\tiny Emu}}) \rbrace_{m}] \neq \mathbb{E}[\lbrace \Omega(\bm{x}_{\text{\tiny Sim}} \rbrace_{n})]$. Therefore such emulations are likely to bias any subsequent statistical analysis. An analogous argument may be made torward $\text{Var}[\lbrace \Omega(\cdot) \rbrace_{m}]$ and other higher order descriptors.

The solution we propose is to instead work with a small \emph{training} ensemble of simulated observables, which more adequately represent the data-generating distribution. During subsequent statistical analysis whenever observables are required, a random latent representation is uniformly drawn from this training set and used to generate $\bm{x}_{\text{\tiny Emu}}$ through the method outlined in Section \ref{sec:approximating_manifold}.

\begin{figure}[t]
  \begin{algorithm}[H]
    \setstretch{1.15}
    \caption{Emulation of cosmic string signatures} \label{alg:emulation_algo}
    \begin{algorithmic}[0]
      \State \emph{First take our small set of $m$ simulations $\lbrace \bm{x}_{\text{\tiny Sim}} \rbrace_{m}$ and compute their latent vectors, which we will condition on during synthesis.}
      \vspace{5pt}
      \Procedure{\small Generate Latent Ensemble}{$\Phi, \lbrace \bm{x}_{\text{\tiny Sim}}\rbrace_m$}
      \For{$i \in [0,m)$}
      \State $\bm{z}_{\text{\tiny Sim}} = \Phi (\bm{x}_{\text{\tiny Sim}})$
      \EndFor
      \State \textbf{return} $\lbrace \bm{z}_{\text{\tiny Sim}} \rbrace_m$
      \EndProcedure
      \vspace{5pt}
      \State \emph{Draw a uniform random reference latent vector $\bm{z}_{\text{\tiny Sim}}$ upon which we will condition. Starting from white noise, use automatic differentiation to find $\bm{x}_{\text{\tiny Emu}}$ such that $\Phi(\bm{x}_{\text{\tiny Emu}}) \approx \bm{z}_{\text{\tiny Sim}}$.}
      \vspace{5pt}
      \Procedure{\small Emulate Field}{$\mu, \Phi, \lbrace \bm{z}_{\text{\tiny Sim}} \rbrace_m$}
      \State $\bm{x} \sim \mathcal{N}(0,\mathbb{I}_{d_{\mathcal{X}}})$ \Comment{Random initial field}
      \State $\bm{z}_{\text{\tiny Sim}} = \lbrace \bm{z}_{\text{\tiny Sim}} \rbrace_{j \sim \mathcal{U}\lbrace 0,m-1\rbrace}$ \Comment{Draw latent vector}

      \Function{\small Loss Function}{$\Phi, \bm{x}, \bm{z}_{\text{\tiny Sim}}$}
      \State $\mathcal{L} = \norm{\Phi(\bm{x}) - \bm{z}_{\text{\tiny Sim}}}_2^2$
      \State \textbf{return} $\mathcal{L}, \frac{\partial \mathcal{L}}{\partial \bm{x}}$ \Comment{Return loss and gradient.}
      \EndFunction
      \State $\bm{x}_{\text{\tiny Emu}} =$ L-BFGS-B($\text{Loss Function}, \bm{x}, \bm{z}_{\text{\tiny Sim}})$
      \State \textbf{return} $\gmu\bm{x}_{\text{\tiny Emu}}$ \Comment{Rescale to string tension $\mu$}
      \EndProcedure
    \end{algorithmic}
  \end{algorithm}
  \vspace{-18pt}
\end{figure}

In this way one may reasonably expect to find that the statistics computed from a set of emulated observables should match those computed on a set of simulated observables. That is to say that $\mathbb{E}[\lbrace \Omega(\bm{x}_{\text{\tiny Emu}})\rbrace_{m}] \approx \mathbb{E}[\lbrace \Omega(\bm{x}_{\text{\tiny Sim}}) \rbrace_{n}]$, provided $n$ and $m$ are each sufficiently large. Increasing the amount of training data will improve the reliability and accuracy with which the distribution of our limited ensemble of training simulations matches the underlying data-generating distribution, improving the degree to which emulated observables are approximately drawn from the true data generating distribution $\bm{x}_{\text{\tiny Emu}} \sim \mathbb{P}_{\text{\tiny True}}$.

To summarise this approach makes the following assertion: the distribution over observables upon which we condition during emulation is transferred to the distribution of our emulated observables. Some augmentation is applied to this distribution, as there is some variability during synthesis, but typically this is a comparatively small effect. Hence, using a small training set of simulated observables provides a straightforward means by which the distribution of emulated observables can be made substantially more realistic.

\vspace{6pt}
\noindent \textbf{Emulation as Augmentation:}~\textit{Our approach may be considered extreme data augmentation, wherein latent emulation bridges the gap between the number of simulations necessary for inference and those which may feasibly be generated. The limited span of our small ensemble of simulations is enhanced by the variability (expressivity) induced by the surjectivity of $\Phi$.}
\vspace{6pt}

Alternatively, one may attempt to enhance the variability of synthetic observables by modelling a probability measure on the latent representation directly, as was promoted by \citet{loaiza-ganem:2022}. In the case where $\Phi$ is a generalized autoencoder the compressive mapping is injective and learned. However, when $\Phi$ is given by the WPHs it is not at all obvious which distribution over latent variables corresponds with $\mathbb{P}_{\text{\tiny True}}$. There are several approaches one may wish to consider however we leave this for future work \citep[see \emph{e.g.}][]{de:2022}.

\subsection{Algorithm and Computational Efficiency}
Our approach involves three primary steps: (1) A small training set of latent vectors is computed from simulations once; (2) A random latent vector $\bm{z}_{\text{\tiny Sim}}$ is drawn from this ensemble; and (3) the loss discussed in Section \ref{sec:approximating_manifold} is minimised to generate an emulated observable such that $\Phi (\bm{x}_{\text{\tiny Emu}}) \approx \bm{z}_{\text{\tiny Sim}}$. These steps are outlined in Algorithm \ref{alg:emulation_algo} and Figure \ref{fig:synthesis}, and are implemented in code which we make publicly available.~\github{https://github.com/astro-informatics/stringgen}\footnote{\url{https://github.com/astro-informatics/stringgen}}

We benchmarked the computational overhead for our approach on a single dedicated NVIDIA A100 Tensor Core GPU with 40GB of device memory. Compiling the PyWPH kernel, our compression $\Phi$, for $1024\times1024$ images takes $\sim$ 11s on average and occupies $\sim$ 27GB of the available onboard memory; indicating the PyWPH software is fast but not yet memory efficient. It should be noted that we adopt default configuration of all PyWPHv1.0 hyper-parameters, and that subsequent PyWPH releases demonstrate further acceleration. Synthesis of a single string induced anisotropies takes $100$ L-BFGS iterations with a wall-time of $\mathcal{O}(100\text{s})$. In practice, the quality of synthetic observations degrades only slightly if the optimiser is run for significantly fewer iterations, and so the wall-time can easily be reduced to less than a minute. As a baseline; a single flat-sky Nambu-Goto simulation at this resolution takes more than a day of wall-time, and a full-sky simulation can take in excess of 800,000 CPU hours.

\section{Validation Experiments}
\label{sec:numerical_results}
To demonstrate the efficacy of the emulation process discussed at length in Section \ref{sec:fast_emulation} and summarised in Algorithm \ref{alg:emulation_algo}, we generate a set of synthetic cosmic string induced CMB anisotropies, the summary statistics of which are validated against those computed over an ensemble of state-of-the-art Nambu-Goto string simulations.

\begin{figure*}
  \centering
  \includegraphics[width=\columnwidth]{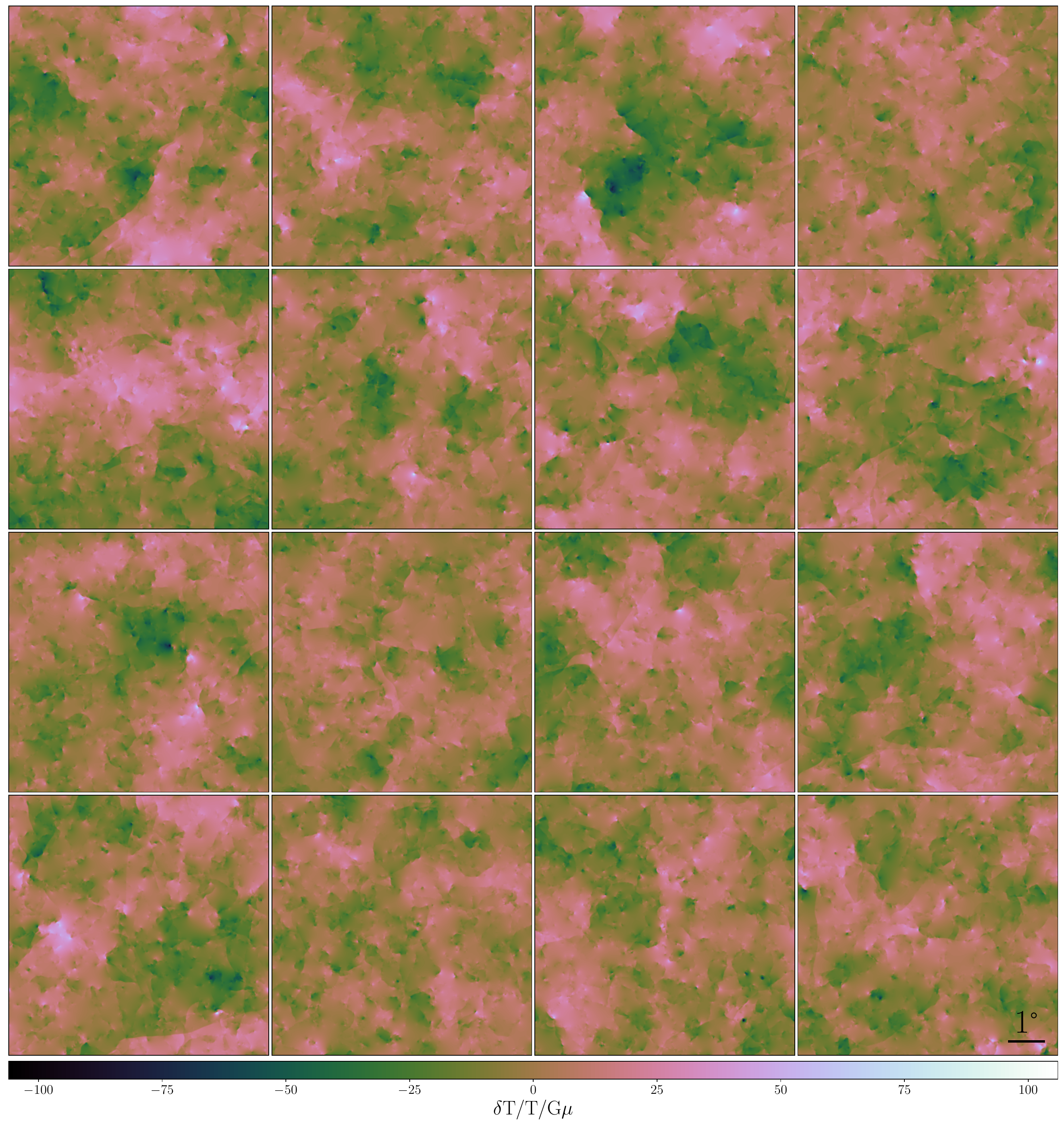}
  \includegraphics[width=\columnwidth]{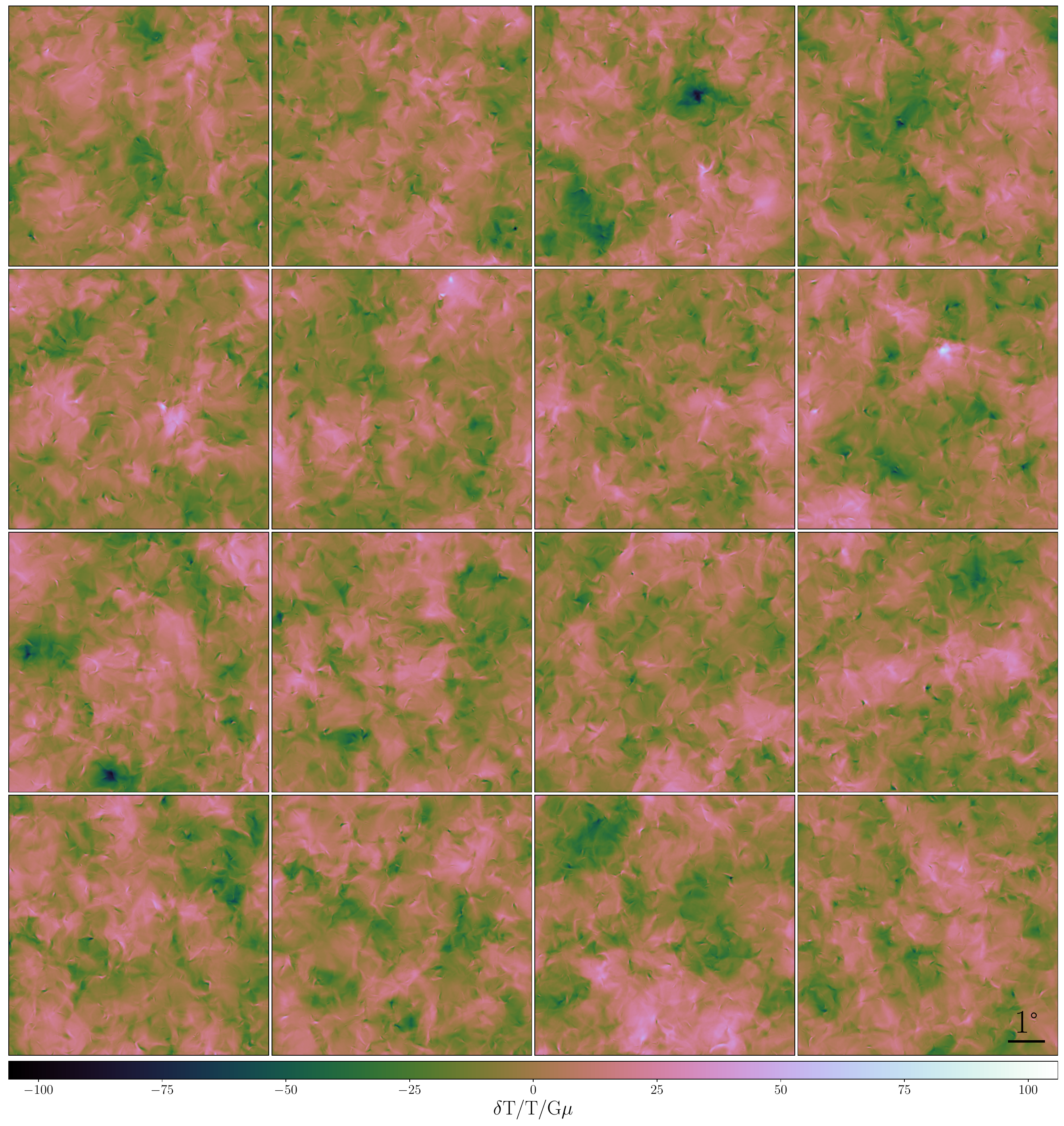}
  \put(-395,-10){Simulations}
  \put(-147,-10){Emulations}
  \vspace{3pt}
  \caption{\textbf{Left:} A gallery of simulated Nambu-Goto cosmic string induced CMB anisotropies randomly sampled from an ensemble of 1000 of such images. Each of these simulations can take in excess of a day to compute. \textbf{Right:} A gallery of emulated string induced anisotropies, each of which take on average under a minute to generate, and are statistically indistinguishable from their simulated counterparts displayed on the left. These synthetic string induced anisotropies are emulated using the methods presented in this article.}
  \label{fig:gallery}
  \vspace{5pt}
  \hrule
\end{figure*}

\subsection{Nambu-Goto String Simulations}
Due to the multiscale nature of wavelets, string induced CMB anisotropies may be emulated for a wide variety of string models, given a field simulation. In this analysis we adopt the Nambu-Goto string simulations of \citet{fraisse:2007}, although in principle alternative string simulations could be considered.

These Nambu-Goto string induced anisotropies are convolved with a 1 arcminute observational beam in line with current ground based observations, such as the Atacama Cosmology Telescope \citep{louis:2014:atacama} and South Pole Telescope \citep{chown:2018:spt}. It is important to note that these simulated flat-sky maps are generated using discrete Fourier transforms and that the genuine cosmic string power spectrum goes as $\sim 1/k$. Therefore, these simulations introduce substantial aliasing at small scales. Such beam convolutions mitigate aliasing by removing any excess power at high frequencies. In total, we have 1000 state-of-the-art Nambu-Goto string maps, each of dimension $1024 \times 1024$, covering a $7.202^{\circ}$ field of view at sub-arcminute resolution.

\subsection{Methodology} \label{sec:methods}
We partition the 1000 available $1024\times 1024$ Nambu-Goto string simulations into \emph{training} and \emph{validation} datasets, with 300 and 700 simulations respectively. For each simulation we compute the associated WPH representation, which we store for subsequent use. Note that we adopt the machine learning nomenclature for consistency, though training is not necessary since we adopt WPHs as our compression $\Phi$, which provide a designed rather than learned latent representation space. Following the method outlined in Algorithm \ref{alg:emulation_algo}, we generate 700 emulated string induced anisotropies, each time uniformly randomly sampling a set of WPH coefficients $\bm{z}_{\text{\tiny Sim}}$ from the training set. Finally, we compute summary statistics over our emulated CMB anisotropies, which we validate against those computed over the validation dataset.

\begin{figure*}
  \centering
  \includegraphics[width=0.645\linewidth]{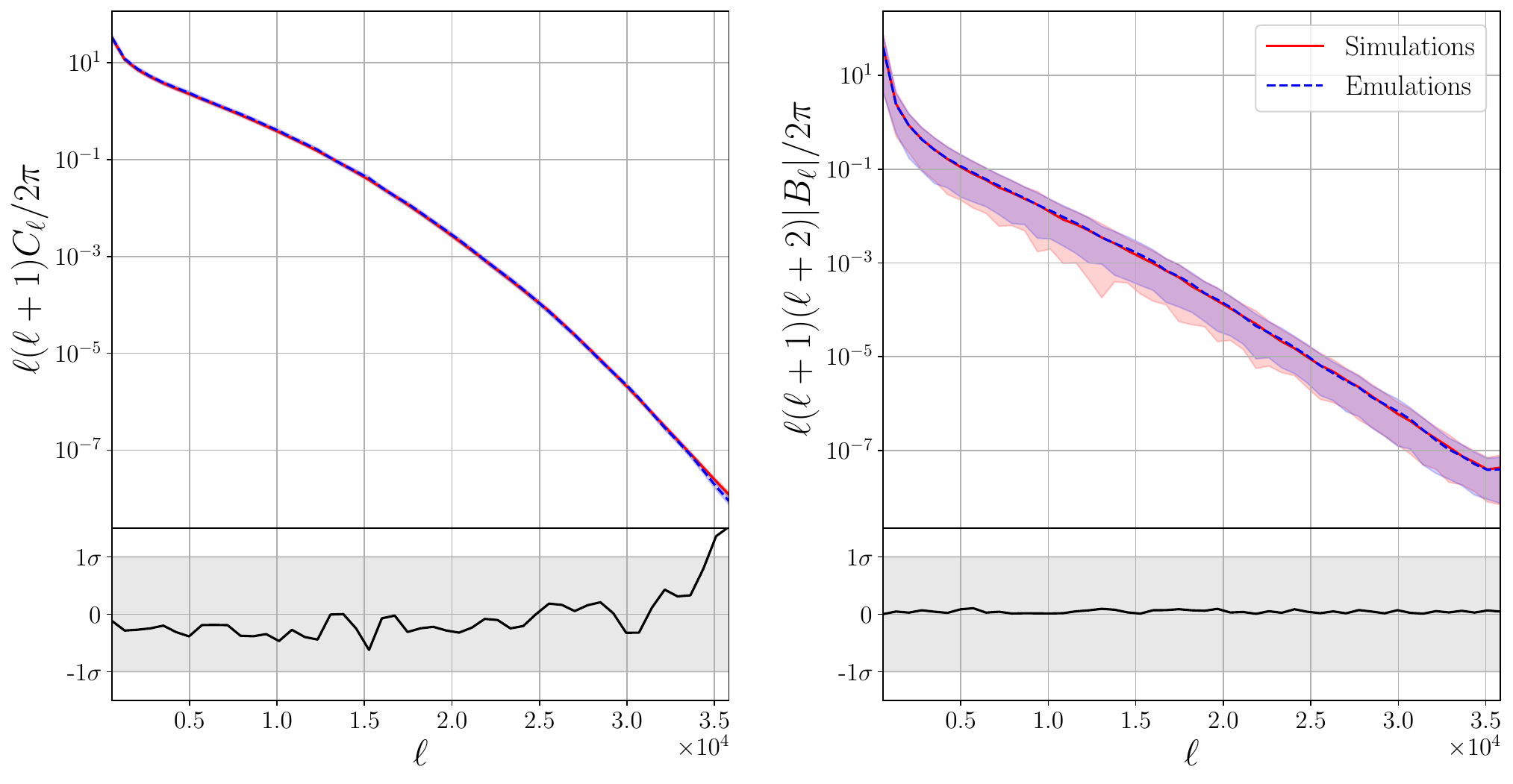}
  \includegraphics[width=0.34\linewidth]{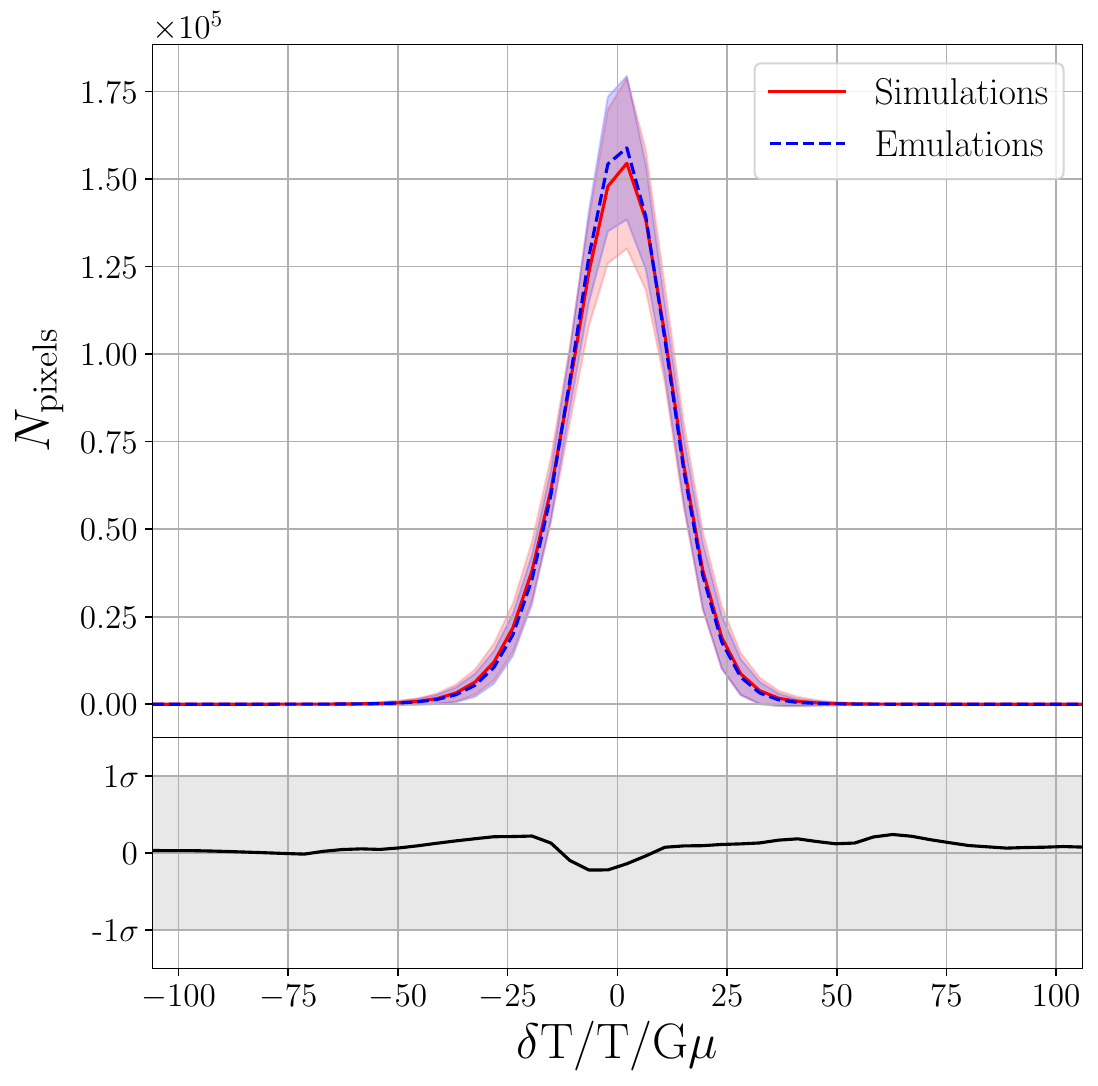}
  \put(-488,3){(a)}
  \put(-317,3){(b)}
  \put(-145,3){(c)}
  \put(-235.4,144.75){\includegraphics[width=0.1055\linewidth, trim={20.4cm 6.5cm 8.3cm 8.8cm}, clip]{figures/spectra_test_em.pdf}}
  \\
  \includegraphics[width=\linewidth]{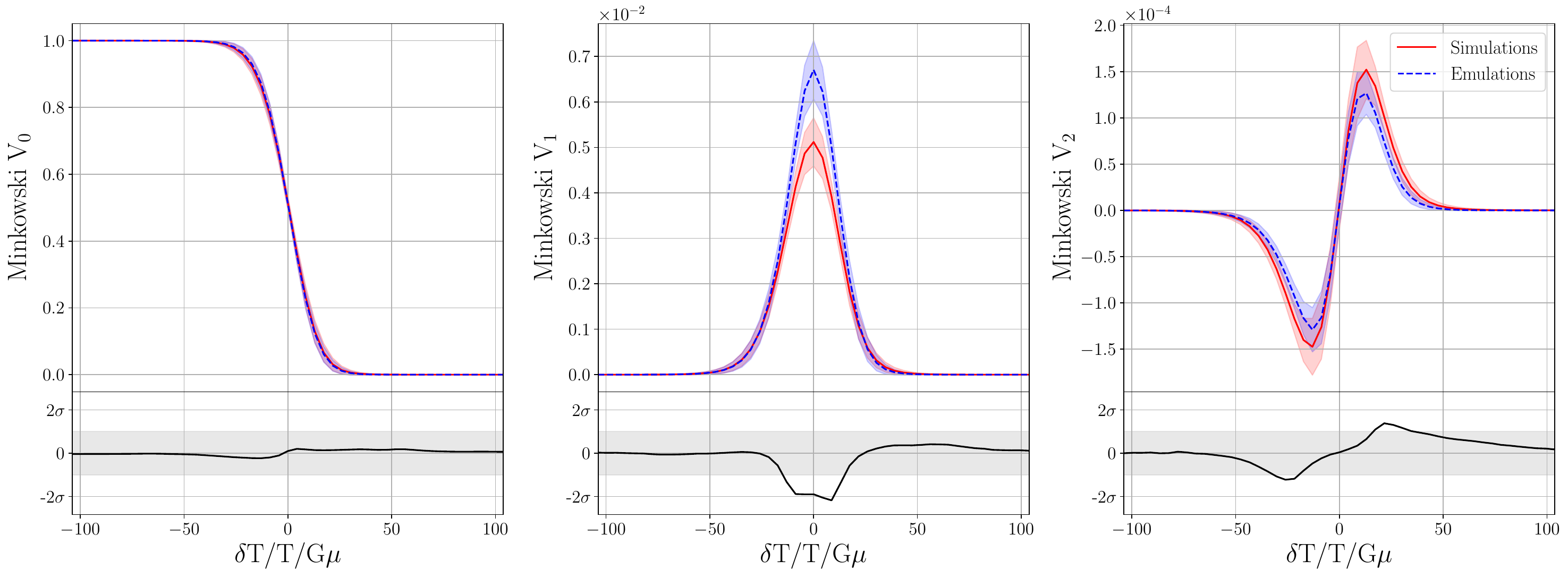}
  \put(-488,3){(d)}
  \put(-317,3){(e)}
  \put(-145,3){(f)}
  \vspace{-3pt}
  \caption{Summary statistics considered for the validation of the emulation techniques presented in this article. Each panel displays the mean of a summary statistic (line) and its variance ($1\sigma$, shaded), for 700 simulated (red, solid) and emulated (blue, dashed) string induced CMB anisotropies. At the bottom of each plot the difference between simulated and emulated anisotropies, in units of $\sigma$, is presented. \textbf{(a)} Standard power-spectrum, for which simulated and emulated statitsics are consistent. \textbf{(b)} Bispectrum, with a flattented triangle configuration, for which both simulation and emulation are statistically indistinguishable. \textbf{(c)} Histogram of pixel intensitites, which are again extremely consistent. \textbf{(d-f)} These sub-figures display the three Minkowski functionals which are, from left to right, sensitive to the area, boundary, and Euler characteristic respectively. For both $V_0$ and $V_2$ simulation and emulation are highly consitent. However, a $\sim 2 \sigma$ discrepency can be seen for $V_1$ around $\delta T/T/\gmu \approx 0$. This exaggerated peak around $0$ is likely due to low-intensity oscillations introduced from the extended support of bump steerable wavelets adopted in the PyWPH package. This effect could be mitigated by the use of alternative wavelets that are better localized in the spatial domain (which is beyond the scope of the current work). Nevertheless these summary statistics are overall in very good agreement.}
  \label{fig:minkowski_and_spectra}
\end{figure*}

\subsection{Validation} \label{sec:validation}
A gallery of randomly selected simulated and emulated string induced anisotropies can be seen in Figure \ref{fig:gallery}; the statistical properties of these maps appear very similar to the eye.
Though it is necessary that emulated observables $\bm{x}_{\text{\tiny Emu}}$ are of high fidelity, one must further ensure that an ensemble of such observables correctly characterise authentic CMB anisotropies. That is, emulated observables for scientific applications must be both of high fidelity and appropriate variability. This duality is discussed in Section \ref{sec:generative_modelling}. One must ensure that $\bm{x}_{\text{\tiny Emu}}$ are, at least approximately, distributed according to the data generating distribution $\mathbb{P}_{\text{\tiny True}}$. If this second condition is not satisfied, although one may recover individual maps which appear reasonable, the aggregate statistics of such maps will likely be incorrect.

Two na\"ive approaches can help elucidate this point. Suppose one selects a single latent vector $\bm{z}_{\text{\tiny Sim}}$ from which many synthetic observables are generated. We explored this and indeed find that the statistics of these anisotropies highly concentrate around the point statistics associated with our chosen latent vector $\bm{z}_{\text{\tiny Sim}}$ and do not fully capture $\mathbb{P}_{\text{\tiny True}}$. Suppose instead one attempts to ameliorate this by constructing an averaged latent representation $\mathbb{E}[\lbrace \bm{z}_{\text{\tiny Sim}} \rbrace_{k}]$ over $k$ training simulations, from which many synthetic observables are generated. Again, we explored this and find that the statistics highly concentrated around the mean latent vector and do not remotely capture $\mathbb{P}_{\text{\tiny True}}$. However, it should be noted that cosmic string induced anistropies exhibit structure which is particularly difficult to model, so it may be that such approaches are sufficient for other applications.

To ensure we capture $\mathbb{P}_{\text{\tiny True}}$ sufficiently to support the use of synthesised observations for scientific inference, we adopt the method outlined in Section \ref{sec:fast_emulation}. We validate these synthetic cosmic string induced CMB anisotropies on a range of popular summary statistics that are senstive to both Gaussian and non-Gaussian information content. Specifically we consider the power spectrum \citep{lizarraga:2014,lizarraga:2014b,lizarraga:2016,charnock:2016}, squeezed bispectrum \citep{planck2013-p20,regan:2015}, Minkowski functionals \citep{gott:1990}, and higher order statistical moments.

Looking to Figure \ref{fig:minkowski_and_spectra}, the power spectrum (Figure \ref{fig:minkowski_and_spectra}a), bispectrum with flattened triangle configuration $B(k,k,k/2)$ (Figure \ref{fig:minkowski_and_spectra}b), and the distribution of pixel intensities (Figure \ref{fig:minkowski_and_spectra}c) are matched to well within $1\sigma$ (grey region). The variance of these statistics accurately mirrors those computed on simulations indicating a similar degree of variability, which is encouraging.

The Minkowski functionals \citep{mecke:1993} of a $d$-dimensional space are a set of $d+1$ functions that describe the morphological features of random fields. For 2-dimensional cosmic string maps $d=2$ and hence there exist three Minkowski functionals $V_{0,1,2}$ which are sensitive to the area, boundary, and Euler characteristic of the excursion set respectively (an excursion set is simply the sub-set of pixels which are above some threshold magnitude). Looking again to Figure \ref{fig:minkowski_and_spectra}, we can see that $V_0$ is recovered near perfectly (Figure \ref{fig:minkowski_and_spectra}d) and $V_2$ is recovered to $\sim1\sigma$ (Figure \ref{fig:minkowski_and_spectra}f), however $V_1$ is accurate away from $\delta T/T/\gmu \approx 0$ but exhibits a $\sim 2\sigma$ difference for $\delta T/T/\gmu \approx 0$ (Figure \ref{fig:minkowski_and_spectra}e).
Given that bump steerable wavelets do not have compact support in pixel-space \citep{allys:2020:new}, which can induce low-intensity extended oscillations, it is unsurprising that the error in $V_1$ is largest around $\delta T/T/\gmu \approx 0$. An alternative family of wavelets could be considered with more compact support or, as mentioned in Section \ref{sec:approximating_manifold}, total variation regularisation could be imposed to induce an inductive bias against such low-intensity oscillations. In fact, precisely such wavelet dictionaries have been developed on the sphere \citep{baldi:2009, mcewen:s2let_localisation}, however we leave exploration in this direction to future work.

\begin{figure}[t!]
  \centering
  \includegraphics[width=\columnwidth]{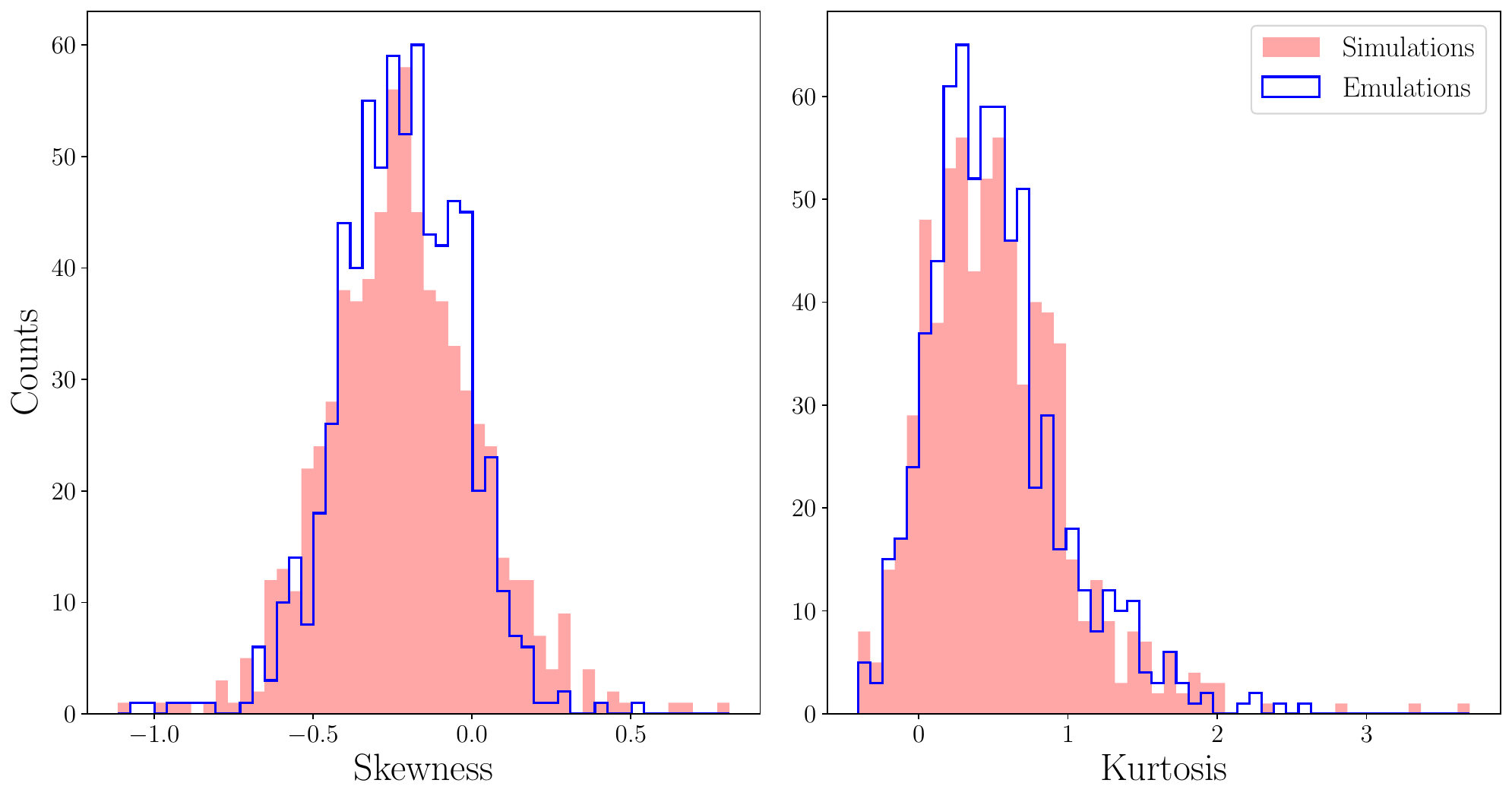}
  \caption{Histograms of the skewness and kurtosis respectively, generated from 700 instances of simulated and emulated Nambu-Goto cosmic string induced CMB anisotropies. We find an excellent agreement between the sets of emulated and simulated cosmic string induced CMB anisotropies.}
  \label{fig:moments}
  \vspace{5pt}
  \hrule
\end{figure}

Finally, in Figure \ref{fig:moments} we consider a histogram of recovered skewness and kurtosis. It should be noted that the kurtosis in particular can be difficult to match, due to a high sensitivity to the tails of a distribution, which are often difficult to capture sufficiently \citep[see \emph{e.g.}][]{feeney:inpainting_isotropy}. Nevertheless, we capture the distribution of both the skewness and kurtosis well.

In summary, although we find a moderate discrepency for one statistic (the second Minkowski functional) around a single threshold (which could likely be mitigated in future by adopting different wavelets in the WPH representation, or subsequent evolutions thereof), all other statistics are excellently matched, both in terms of bias and relative variability.

\vspace{20pt}
\section{Conclusions}
\label{sec:conclusions}
In this article we consider generative modelling, highlighting the differences between its application to natural images and for physics. In contrast to typical use-cases for natural images, in physics it is important to not only generate realistic emulations but to also faithfully trace the underlying probability distribution of fields. We ground this discussion within the context of cosmic string induced CMB anisotropies, which are structurally complex and highly computationally expensive to simulate. For scientific applications, generative models must not only generate realistic observables, but also ensure these synthetic observables are correctly distributed; a qualification which is often overlooked.

Leveraging the recently developed wavelet phase harmonics as a compressed latent representation, we present a method by which cosmic string induced anisotropies may accurately be synthesised at high-resolutions in under a minute. For context, flat-sky string simulations typically take more than a day to evolve, and full-sky simulations take in excess of 800,000 CPU hours. Importantly, our method requires significantly less data, which is a fundamental barrier for the application of many generative modelling techniques to cosmology. Our synthetic observations are statistically commensurate with those from simulated observations. In the spirit of reproducibility and accessibility our code has been made publicly available \github{https://github.com/astro-informatics/stringgen}.

Throughout, we consider the case where strings are generated from a Nambu-Goto action, however in principle the techniques we develop may equally be applied to other string models. For example, one may also emulate anisotropies induced by more complex scenarios such as cosmic superstring networks \citep[\textit{e.g.}][]{urrestilla:2008}. To accommodate fields with increased complexity, more expressive third generation scattering representations are likely to be useful \citep[\textit{e.g.}][]{cheng:2023:scattering}.

Although this work highlights the exciting potential for fast emulation of cosmic string induced CMB anisotropies, it is currently limited to the flat-sky. For wide-field observations (\emph{e.g.} Planck) the sky curvature inevitably becomes non-negligible, hence the extension of these generative modelling techniques to the sphere is neccesary. First generation wavelet scattering techniques on the sphere were developed in previous work \citep{mcewen:scattering}. In ongoing work we are developing accelerated and automatically differentiable spherical harmonic (Price \emph{et. al.} 2023a in prep), wavelet transforms (Price \emph{et. al.} 2023b in prep) and third generation spherical scattering covariances (Mousset \emph{et. al.} 2023 in prep). Note that such third generation scattering covariances have already shown much promise over flat spaces \citep{cheng:2023:scattering}. We are also exploring the fusion of these emulation techniques with simulation based inference, for application to many open areas of astrophysics.


\pagebreak

\section*{Acknowledgements}
The authors would like to thank Dipak Munshi, Tom Kitching, and Luke Pratley for discussion in the early stages of this work. Furthermore we would like to thank Erwan Allys for advice on wavelet phase harmonics, and both Christophe Ringeval and François Bouchet for providing the Nambu-Goto simulations, which are featured throughought this article.
The cosmic string simulations have been performed thanks to computing support pvided by the Institut du D{\'e}veloppement des Ressources en Informatique Scientifique and the Planck-HFI processing center at the Institut d'Astrophysique de Paris. This work used computing equipment funded by the Research Capital Investment Fund (RCIF) provided by UKRI, and partially funded by the UCL Cosmoparticle Initiative. MAP and JDM are supported by EPSRC (grant number EP/W007673/1). MM and AM are supported by the STFC UCL Centre for Doctoral Training in Data Intensive Science (grant number ST/P006736/1). ASM is supported by the MSSL STFC Consolidated Grant (grant number ST/W001136/1) and the Leverhulme Trust.

\section*{Contribution Statement}

Author contributions are specified below, following the Contributor Roles Taxonomy (CRediT).

MAP: Conceptualisation, Methodology, Software, Validation, Investigation, Supervision, Writing (Original Draft, Review \& Editing);

MM: Methodology, Software, Validation, Investigation;

MMD: Validation, Data Curation, Visualisation;

ASM: Conceptualisation, Validation;

AM: Software, Data Curation;

JDM: Conceptualisation, Methodology, Validation, Supervision, Writing (Original Draft, Review \& Editing).

\bibliographystyle{mymnras_eprint}
\bibliography{bib_journal_names_long,bib_myname,bib,mybibs_new}

\end{document}